\magnification 1200
\def\blankline{\par\vskip 12 pt\noindent}
\def\ref{\par\noindent\hangindent 20pt}
\parindent 15pt
~
\vskip 2truecm
\centerline{\bf HEAVY ELEMENT DIFFUSION AND GLOBULAR CLUSTER AGES.} 
\vskip 2truecm
\centerline{V.Castellani$^{1,2}$\footnote*{Send offprint requests to
V. Castellani Dipartimento di Fisica,
Universit\`a di Pisa, I-56100 Italy.}, F.Ciacio$^3$, S.Degl'Innocenti$^3$
     \& G.Fiorentini$^3$.}
\vskip 1truecm
\centerline{$^1$Dipartimento di Fisica, Universit\`a di Pisa, I-56100 Pisa.}
\centerline{$^2$Osservatorio Astronomico Collurania, I-64100 Teramo.}
\centerline{$^3$Dipartimento di Fisica Universit\`a di Ferrara}
\centerline{and INFN, Sezione di Ferrara, I-44100 Italy.}

\vskip 3truecm
\centerline{Running head: Castellani et al.: Diffusion and cluster ages}
\centerline{Main Journal, Section 6.}
\centerline{Thesaurus codes: 02.04.2, 08.05.3, 08.07.1}

\vskip 5truecm
\centerline{Send proofs to: V. Castellani Dipartimento di Fisica,
Universit\`a di Pisa, I-56100 Italy}

\vskip 6truecm
\noindent
\blankline
\blankline
\eject
\baselineskip=24pt
\noindent
{\bf  Abstract}
\par
This paper discusses low mass, metal poor stars, presenting new 
evolutionary computations which take into account
the inward diffusion of both He and heavy elements during the H
burning phase.
The investigation has been extended to He burning models,  
discussing the luminosity of the Zero Age Horizontal Branch 
(ZAHB) models originated from progenitors with efficient element diffusion.
We find that diffusion of heavy elements plays a non-negligible role
in the evolutionary paths of the model.
One finds that element diffusion has a sensitive
influence on evolutionary models, a less significant effect
on the isochrone turn off and, finally, a rather marginal 
impact on the difference in luminosity between the ZAHB and the 
isochrone Turn-Off (TO). The consequences of these results on 
current evaluations of globular cluster ages are discussed
in connection with possible sources of errors.

\par


\blankline

{\bf 1. Introduction} 
\par
\blankline
Over the last years the well known problem of solar neutrinos has stimulated
an increasing attention toward theoretical predictions concerning
solar structure.
Accordingly, several Standard Solar Models have been presented in the
literature, following the continuous improvement in the
available input physics.
However, in the meantime the progress of 
helioseismologic investigations has produced rather severe constraints on
the inner solar structure, bringing to light a non-negligible 
discrepancy between the actual Sun and all
solar models as evaluated according to the traditional
evolutionary scenario. As a matter of fact, one finds that all these
models underestimate the depth of the subatmospheric convection which
affects solar structure as, more in general, it affects the structure
of all other low mass stars.

This troublesome discrepancy has been recently interpreted as an evidence
that the diffusion of He and heavy
elements plays a not marginal role in solar evolution. 
The efficiency of element diffusion in stellar structure has been
a debated problem. Deliyannis \& Demarque (1991) presented arguments
for the existence of mixing mechanisms limiting - at least in the
surface layers - such efficiency. However the ``signature'' of element diffusion
has become progressively evident in the results of helioseismology,
as discussed by Cristensen-Dalsgaard (1993), Guzik (1993) and Bahcall \&
Pinsonneault (1995). According to Bahcall et al. (1996) one should conclude
that solar models that not include diffusion, or with significant internal mixing,
are effectively ruled out by helioseismology.
When this mechanism is taken into account,
Solar Standard Models (SSM) reach a beautiful agreement with 
helioseismology constraints.
According to such an evidence, our FRANEC evolutionary code has been 
recently implemented to account for element diffusion following the 
prescriptions given by Bahcall \& Pinsonneault (1995).
As expected, one finds in this way a SSM which nicely overlaps
the results of the two above quoted authors
(Ciacio, Degl'Innocenti \& Ricci 1996).

Since the evolutionary path of the Sun appears to be deeply affected
by element diffusion, it is of obvious
interest to extend the investigation to the field of old metal poor
Population II stars.
One expects in this way to shed new light on the debated problem
of globular cluster ages and related constraints
on the age of the Universe.
As a matter of fact, one cannot disagree with
Proffitt \& VandenBerg (1991) 
when they state that similar constraints 
``can be taken seriously only if ALL of the important physics''
is correctly taken into account.

He diffusion in low mass stars has already been discussed
in the above quoted exemplary paper by Proffitt \& VandenBerg
(1991: hereinafter referred to as PVB)
who presented and discussed the effect of He diffusion on 
globular cluster isochrones.
In their conclusions, these authors pointed
out the possible relevance of significant gravitational settling
of elements heavier than He. However, further investigations on
the matter (Chaboyer, Sarajedini \& Demarque 1992, Chaboyer 1995,
Canuto et al. 1996)  keep considering the sedimentation of He only. 
On the other hand, sedimentation of heavy elements has a non-negligible 
influence on solar standard models, and we will show in the following
that this is also true for metal poor stars of low masses. 
   
In this paper we will study the consequences of introducing 
He and heavy elements diffusion into theoretical evolutionary 
structures for low mass, metal poor stars. In the next section 
we will discuss the evolutionary behaviour of selected stellar models, 
as computed under different assumptions about the efficiency of 
diffusion to elucidate the contribution of various mechanisms 
concerning element sedimentation.
The evolutionary computations presented are in all cases based on the
most updated input physics used for producing the SSM, as
described in Ciacio et al. (1996).
Present stellar models
without diffusion can thus be regarded as an update of similar 
models already presented in the literature 
(Straniero \& Chieffi 1991, Castellani, Chieffi \& Pulone 1991).
                                             
In section 3 we will present new cluster isochrones,
discussing their influence on the present debate about cluster ages.
In the same section we will extend the investigation to He burning models,  
discussing how the luminosity of the Zero Age Horizontal Branch
(ZAHB) is affected by element diffusion in the progenitor models.
As a result, we will find that element diffusion has a sensitive
influence on evolutionary models, a less significant effect
on the isochrone turn off and, finally, a rather negligible
impact on the difference in luminosity between the ZAHB and the 
isochrone Turn-Off (TO). The consequences of these results on 
current evaluations of globular cluster ages will be finally
discussed.

\blankline

{\bf 2. Element diffusion in low mass metal poor stars.} 

To gain insight on the problem of diffusion let us first discuss
the evolutionary behaviour of a stellar model chosen to represent typical
metal poor globular cluster stars in some detail, namely adopting a
mass M= 0.8 M$_{\odot}$ together with original He and metallicity as given 
by Y=0.23 and Z=0.0004, respectively.
Figure 1 shows the evolutionary track of such a model,
as computed under different assumptions about the efficiency of diffusion.
According to labels in that figure one finds: 
i) the traditional track (ND) where diffusion is not taken into account, 
ii) a track (dashed) where only He is allowed to diffuse,
iii) a track (M1) with diffusion of both He and heavy elements, but without 
accounting for the effect of sedimentation of metals on matter opacity and,
iv) our "best" final track (M2) where all diffusion effects are taken into
the right account.

As already known, one finds that He sedimentation makes the evolutionary
track redder, with a lower luminosity of the turn off (TO) which is
reached earlier. This last occurrence is shown in Table 1 where we report, 
among other selected quantities, the age of the various models at the TO.
From data in Table 1 one recognizes that He sedimentation 
makes the track turnoff fainter by more than $\delta$logL=0.08 and 
younger by more than 1 Gyr. These results can be compared with
the results by Proffitt \& Michaud (1991), who obtained for a very similar model
$\delta$logL=0.05 with an age decreased by 0.75 Gyr. It appears that our
approach to diffusion gives similar but larger effects on the
topology and on the timing of the evolutionary track.

Figure 1 and Table 1 show that the diffusion of heavy elements
Šplays a further role only through the contribution to stellar opacity, 
plays a further role only through the contribution to stellar opacity,
partially counteracting the effect of He depletion. As a whole,
when the diffusion of both He and heavy elements is taken into
account one finds a track TO fainter (with respect to the case
without diffusion) by $\Delta$logL=0.064, the TO being reached in
a time shorter by a bit less than 1 Gyr.
It turns out that our results with sedimentation of both He and heavy elements 
approach the results by Proffitt \& Michaud (1991) where only He 
sedimentation was taken into account.
It follows that one can easily
foresee that our isochrones will closely approach the diffusion
scenario already discussed by PVB. 

Figure 2 shows the effect of diffusion on the 
distribution of selected elements throughout the structure
of our 0.8 M$_{\odot}$ model in the phase of exhaustion of central hydrogen,
when surface abundances are reduced to
Y$_{atm}$= 0.152 and Z$_{atm}$=0.0003.
However, the surface abundances of these elements will be restored 
during the first dredge up in the Red Giant phase. 
Eventually, one finds that our model approaches its He flash
with  Y$_{atm}$= 0.228 in the atmosphere against Y$_{atm}$=0.241
expected if diffusion were not at work.
In this respect, one finds that 
the model behaviour is not far from the evolutionary scenario depicted
by PVB, with very similar values for the decrease of surface 
He. According to the discussion already given by PVB, this decrease
implies a decrease in the luminosity of the HB by a small 
but not completely negligible amount. On the contrary, one finds that 
the temporal depletion of heavy elements in the stellar atmosphere 
is eventually almost completely smoothed
away by the dredge up, the final metallicity of the Red Giant 
coming back to Z=0.00038, with an expected minor influence on the
following He burning structures.

Finally, figure 3 compares evolutionary tracks with or without
diffusion for selected choices of the mass of the evolving stars. 
Table 2 gives selected quantities for all these evolutionary models
taken at their track turnoff. Left to right one finds the mass M of
the model (in solar masses), the assumption about the efficiency of 
diffusion, luminosity, temperature and age of the model at the 
track turnoff and the variation both in luminosity ($\Delta$logL)
and in age ($\Delta$t/t) at the TO caused by diffusion.

Inspection of these data discloses that the effects of diffusion 
show a maximum in stars with masses around M=0.8 M$_{\odot}$. 
The irrelevant differences are within the accuracy
of theoretical evolutionary computations. However, one finds that
Proffitt \& Michaud (1991), who integrated structures with masses equal
or below 0.8 M$_{\odot}$, already found that the
0.8 M$_{\odot}$ model is the most affected by diffusion.
An occurrence to be related to
the progressive sinking of external convection when going toward 
less massive cooler stellar structures.
Moreover, the lower efficiency of diffusion in the 0.9 M$_{\odot}$ model
can be understood as a consequence of the shorter evolutionary
times, according to the evidence that diffusion is a slow mechanism
which needs time.
We are thus inclined to regard the quoted maximum as a
feature rather than an artifact of the computations. 

\par
\blankline

{\bf 3. Cluster isochrones and age calibrators.}

In the previous section we discussed the variation in the
luminosity and in the age of the track TO caused by
inward diffusion of elements. One may notice that, 
if isochrone TO were to suffer similar variations, according 
to current calibrations of TO luminosities in
terms of age (as given --e.g.-- in
Castellani, Chieffi \& Pulone 1991), the quoted results would imply 
that diffusion rejuvenates a cluster
by no less than 4 Gyr if only He is depleted, or by about 3 Gyr 
if both He and heavy elements are allowed to diffuse. However, as pointed 
out by Proffitt \& Michaud (1991) and reinforced by PVB, the proper way of 
discussing ages is through cluster isochrones since
consideration of evolutionary tracks on their own exposes one to
the risk of misleading results.

According to a well known procedure, evolutionary tracks can be
interpolated to eventually produce theoretical isochrones
predicting the distribution in the HR diagram of H burning
cluster stars for the adopted chemical composition and for various
assumptions about the cluster age.
Figure 4 shows a set of similar isochrones as computed with or without
diffusion for ages ranging from 10 to 15 billion years, whereas Table 3
gives selected quantities concerning the isochrone turn off.

It appears that over all the explored ranges of ages the effect 
of diffusion on isochrone TO is about one half the effect found 
on track TO's. This reinforces that evolutionary tracks
alone cannot be used to obtain reliable predictions about the behaviour
of isochrones TO's.
We note that the difference between track and isochrone
TO's is a general feature, but enhanced in the diffusive case by the
already presented differential effect of diffusion on the evolutionary 
times of stars of different masses. This is shown in Figure 5 where
we compare the evolutionary track for a 0.8 M$_{\odot}$ model, with 
and without diffusion to the corresponding
isochrones as evaluated for the age when the model is just
reaching its track TO.
As a whole, data in Figure 4 and Table 3 show that,
if diffusion is taken into account,
current evaluation of cluster ages based on the luminosity of
isochrone TO should be
rejuvenated by about 1 Gyr, in agreement with the current evaluation
already given accounting for the inward diffusion of He only.

Computational results concerning H burning Red Giant structures 
can be finally used to evaluate He burning structures as produced
by progenitors where diffusion has been active.
The 0.8 M$_{\odot}$
model with diffusion ignites He within a He core of M$_c$=0.4961 M$_{\odot}$
against the value of M$_c$=0.4993 M$_{\odot}$ in the case without diffusion.
This is not an unexpected result, since when diffusion is
at work a giant reaches He ignition with a
lower He abundance in the envelope, and one knows that
stars with lower original He 
have larger He cores (but lower HB luminosity!).
However, original He and He in
the convective envelope of a giant are not exactly the same 
thing, and the above result indicates that the actual mass of the He core
has been governed by the amount of He in the stellar envelope, indipendently
of the original values (Y$_{or}$) of Y.

As a matter of fact, from Bono et al. (1995)
in the interval Y$_{or}$=0.20$\div$0.23 one derives
dM$_c$/dY$_s$=$-$0.14 (dM$_c$/dY$_{or}$=$-$0.12)
and taking $\Delta$ Y$_s$=0.013
from the difference in the atmospheric He after the first dredge up,
one finally obtains $\Delta$M$_c$=0.0018, i.e.,
exactly the difference found in direct computations
with and without diffusion.
Note that this is not an obvious result.
As a matter of fact, numerical experiments performed artificially decreasing
Y$_s$ in the atmosphere of a giant in the upper portion of the RG branch
shows that M$_c$ is unaffected by such a variation.
This result can be understood by recalling that mass loss
in a red giant does not affect the
growth of the central He core for the simple reason that 
the thermodynamic time scale of the core is larger than the evolutionary
times in the red giant phase (Castellani \& Castellani 1993).
For the same reason, changes of He abundances in the envelope 
which surrounds a luminous evolving giant do
not affect the evolution of the core.
On the contrary, the quoted evolutionary result indicates that the depletion
of He in the stellar envelope resulting from the combined effects
of diffusion and dredge up occurs early enough to allow the stellar core to be
sensitive to this variation.

Thus Zero Age Horizontal Branch models with diffusion are expected with 
surface He smaller by $\Delta$Ys=$-$0.013,
and with an He core mass appropriate for this atmospheric He abundance.
By taking again from Bono et al. (1995) dlogL$_{ZAHB}$/dY$_s$=1.7
one can predict that HB luminosity level
(at the temperature of the RR Lyrae gap) for stars which
experienced diffusion should be fainter by about $\Delta$logL=$-$0.022.
Figure 6 shows that such a prediction is fully confirmed by detailed
computations of He burning models.
As a result, one predicts that diffusion 
should have a negligible effect on the calibration
of the difference in luminosity between HB and TO in terms of
the cluster age.
As a matter of fact, the expected increase $\Delta$logL$\sim$0.01 for
each observed $\Delta$logL$_{TO}^{ZAHB}$
implies a decrease in the estimated age of a few 10$^8$ years.

\blankline

{\bf 4. Discussion and conclusions.} 
\blankline
In this paper we studied the effect of element diffusion
on the evolution of old metal poor stars. We found that inward
diffusion of heavy elements partially counteracts the effects
of He diffusion. As a result, one finds that when diffusion
of all elements is taken into account the luminosity of the 
isochrone TO should decrease of $\Delta$logL$\sim -$0.03.
In terms of calibration of cluster ages, a similar decrease would
imply a rejuvenation of clusters by about 1 Gyr, a bit less but
in substantial agreement with the rejuvenation produced by consideration
of He diffusion only, as discussed in several papers
(see, e.g., Chaboyer 1995 and references therein).

However, as a consequence of diffusion, red giants at the onset 
of the He flash are expected with lower abundance of He in the envelope,
a variation that has to be taken into account
and that goes in the direction of decreasing the luminosity
of HB stars.
Direct computations of He burning HB models
confirm that prediction, producing models less luminous by  
$\Delta$logL$\sim -$0.02.

Comparison with Proffitt \& VandenBerg (1991) show that we find 
smaller decreases of turnoff
luminosity but similar decreases of the HB luminosity, which now
exactly compensates the decrease in luminosity of the turnoff
for a given age.
This of course reinforces the statement by PVB
which already indicated that cluster ages derived from the 
difference in magnitude between HB and TO are much less sensitive to 
uncertainties in the diffusion rates than ages derived from the TO
luminosity only, adopting some other calibrator of the 
cluster distance modulus.    

According to this theoretical evidence, one should conclude the recent
evaluation of about 12 Gyr given for metal poor galactic globulars
(Castellani, Brocato \& Piersimoni 1996)
should safely survive even if element diffusion is taken into account.
However, to assess the proper meaning of the above evaluation let us here
discuss possible sources of errors.
The influence of the input physics and/or of element abundances on the
turn--off luminosity has been recently and carefully investigated in a paper
by Chaboyer (1995) to which we address the interested reader.
Here let us only add some further comments concerning theoretical predictions
about the difference in magnitude between HB and TO,
bearing in mind -- as an order of magnitude --
that decreasing the difference by $\Delta$logL$_{TO}^{ZAHB} \simeq$0.1
means decreasing the evaluation of the cluster age by about 4 Gyr.

The influence of cluster metallicity on both TO and HB luminosity has been
already evaluated by Chaboyer (1995), by including an empirical dependence
of HB luminosity on metals.
With our fully theoretical approach we find dlogL$_{TO}$/dlogZ$\simeq -$0.13
and dlogL$_{HB}$/dlogZ$\simeq -$0.075.
On this basis one finds dlogL$_{TO}^{ZAHB}$/dlogZ$\simeq$0.06
which confirms the minor relevance of a precise determination of cluster
metallicities as far as the cluster ages is concerned.
The influence of the adopted amount of original He is however much more
relevant.
To introduce this problem let us first notice that the usual procedure to
evaluate Y through the number ratio R of HB to Red Giant stars more luminous
than the HB luminosity level is open to serious uncertainties,
particularly when applied to metal poor clusters with poorly populated
instabilities strips.
From RG evolutionary times and R calibrations, as given in Bono et al. (1995),
one finds that an error of 0.2 mag. in the bottom luminosity of the RG sample,
possibly produced by errors in the HB luminosity level AND in the bolometric
correction for RG stars, gives Y values ranging from Y=0.20 to Y=0.26,
which is exactly the range explored by Chaboyer (1995).

By taking from Iben \& Renzini (1984) dlogL$_{TO}$/dlogY$\simeq$-0.42,
over the quoted range of Y one finds a variation
$\Delta$logL$_{TO} \pm$0.013,
which indicates, as reported by Chaboyer (1995), that such a variation
in Y scarcely affects the estimated cluster ages.
However, theory tells us that at the same time the HB luminosity increases
with Y as dlogL$_{HB}$/dlogY$\simeq$1.7.
As a whole one finds $\Delta$logL$_{TO}^{HB} \simeq \pm$0.06,
which in terms of ages means an error of the order of 2 Gyr.
Thus the estimate of Y could play a relevant role in assessing cluster
ages through the observed differences
$\Delta$logL$_{TO}^{HB}$,
the ages decreasing if Y increases.

As a final point, we note that Chaboyer (1995) has already shown that
uncertainties in the current evaluations of nuclear reaction cross sections
scarcely affect the evaluations of the turn--off luminosity.
However, on theoretical grounds one expects that the 3$\alpha$
reactions do govern the He ignition in the RG core, thus determinig 
 the mass of the core at the ignition and, in turn, the luminosity
of a new born HB star.
According to Rolfs \& Rodney (1988), one finds that the cross section for
this reaction is poorly known, with an estimated error of about 15\%
(at 1$\sigma$).
To make this further point clear we performed numerical simulations of RG
evolution artificially moving the adopted cross section by $\pm 30\%$.
As a result, we found that by increasing (decreasing) the cross section by
such an amount the mass of the He core at the flash decreases (increases)
by 0.002(0.003).
Since $\Delta$logL$_{HB}$/dM$_c \simeq$3.4
(from Sweigart \& Gross, 1978)
one may easily see
that  the quoted cross section scarcely affects the evaluations of ages.
As a conclusion, one finds that current evaluations of cluster ages appear
generally well grounded in the theory, with the warning that the amount
of original He is the main ingredient which appears to be able to move
these evaluations by more than 1 Gyr.

\par
\par
\vfill
\eject
\nopagenumbers
\noindent

{\bf References}
\blankline

\ref{Bahcall J.N. \& Pinsonneault M.H. 1995, Rev. Mod. Phys 76,781.}
\ref{Bahcall J.N., Pinsonneault M.H., Basu S. and Christensen-Dalsgaard J.
preprint IASSNS-AST 96/54}
\ref{Bono G., Castellani V., Degl'Innocenti S. \& Pulone L. 1995,
A\&A 297,115.}
\ref{Canuto V., Caloi V., D'Antona F. \& Mazzitelli I., 1996, preprint.}
\ref{Castellani V., Brocato E. \& Piersimoni A. 1996, (in preparation)}
\ref{Castellani M. \& Castellani V. 1993, ApJ 407, 649.}
\ref{Castellani V., Chieffi A. \& Pulone L. 1991, ApJS 76, 911.}
\ref{Chaboyer B. 1995, ApJ 444, L9.}
\ref{Chaboyer B., Sarajedini A. \& Demarque P. 1992, ApJ 394, 515.}
\ref{Ciacio F., Degl'Innocenti S. \& Ricci B. 1996, A\&A (in publication).}
\ref{Christensen-Dalsgaard J., Proffitt C.R. and Thompson M.J. 1993,
ApJ 403, L75}
\ref{Deliyannis C.P. \& Demarque P. 1991, ApJ 379, 216} 
\ref{Guzik J. A. \& Cox A.N. 1993, ApJ 411, 394}
\ref{Iben I.Jr \& Renzini A. 1984, Physics Report 105,329.}
\ref{Proffitt C.R. \& Michaud G. 1991, ApJ 371, 584.}
\ref{Proffitt C.R. \& VandenBerg D.A. 1991, ApJS 77, 473.}
\ref{Rolfs C. \& Rodney W. 1988 ``{\it Cauldrons in the cosmos}'' The
University of Chicago Press, Chicago.}
\ref{Straniero O. \& Chieffi A. 1991, ApJS 76, 525.}
\ref{Sweigart A.V. \& Gross P.G. 1978, ApJS, 36, 405.}

\eject

\centerline{\bf Figure captions}
\blankline \blankline
\ref{Fig. 1. The HR diagram evolution of the 0.8 M$_{\odot}$ as
computed under different assumptions about the efficiency of
diffusion (see text).}
\ref{Fig. 2. The  distribution of selected elements in the
structure of a 0.8 M$_{\odot}$ model with diffusion near the exhaustion
of central H.}
\ref{Fig. 3. Evolutionary tracks with or without diffusion and for
the labeled assumptions about the mass of the evolving star.} 
\ref{Fig. 4. Selected cluster isochrones with or witout diffusion 
and for the labeled assumptions about the cluster age.}
\ref{Fig. 5. The evolutionary track (dashed
line) of the 0.8 M$\odot$ model with or without diffusion
(upper and lower panel respectively)
compared with the isochrone where the model attains
the track TO.}
\ref{Fig. 6. Zero Age Horizontal Branches  for the 
original chemical composition Y=0.23 and Z=0.0004, as originated 
from progenitors with (full line) or without (dahed line) diffusion.}
\vfill
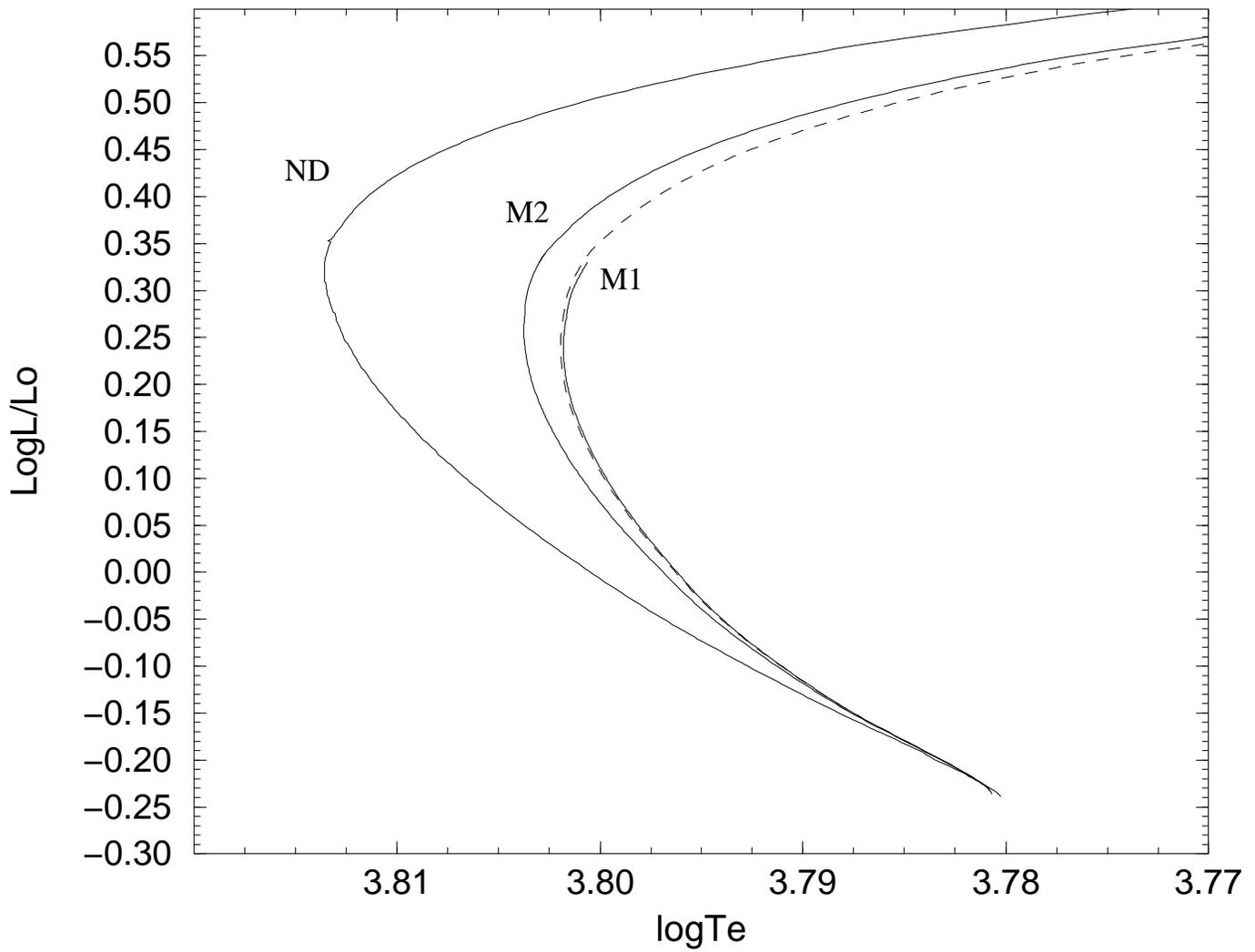
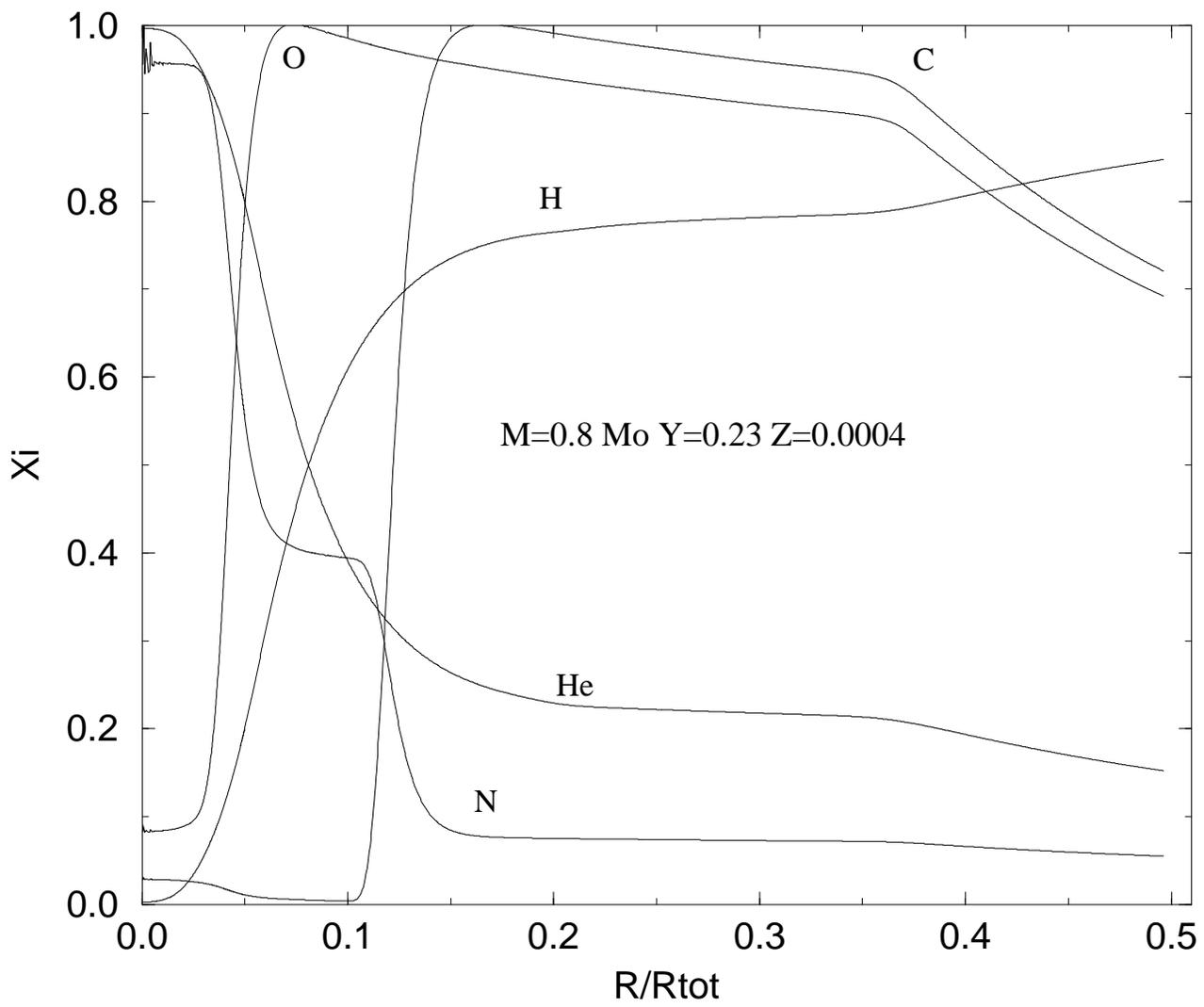
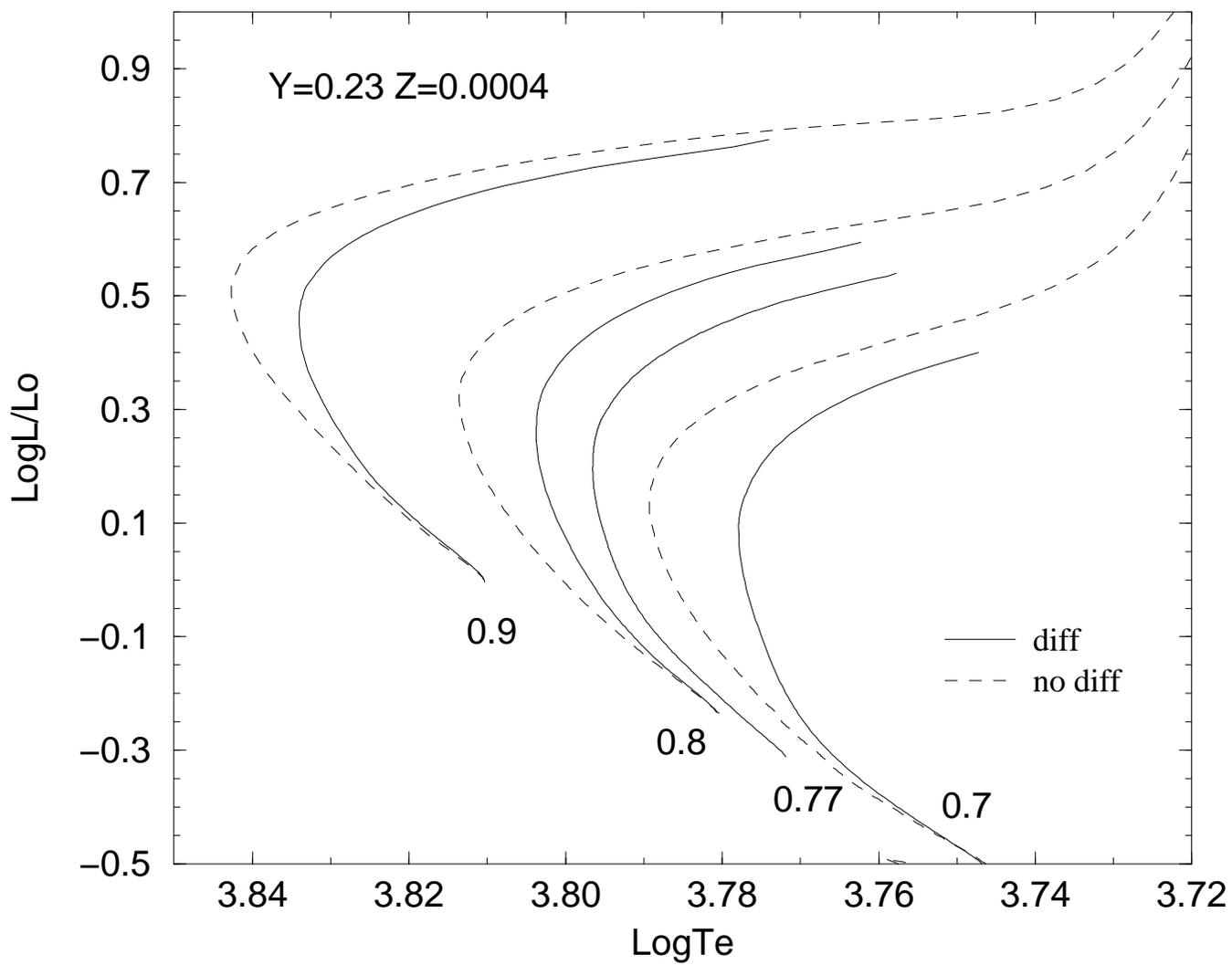
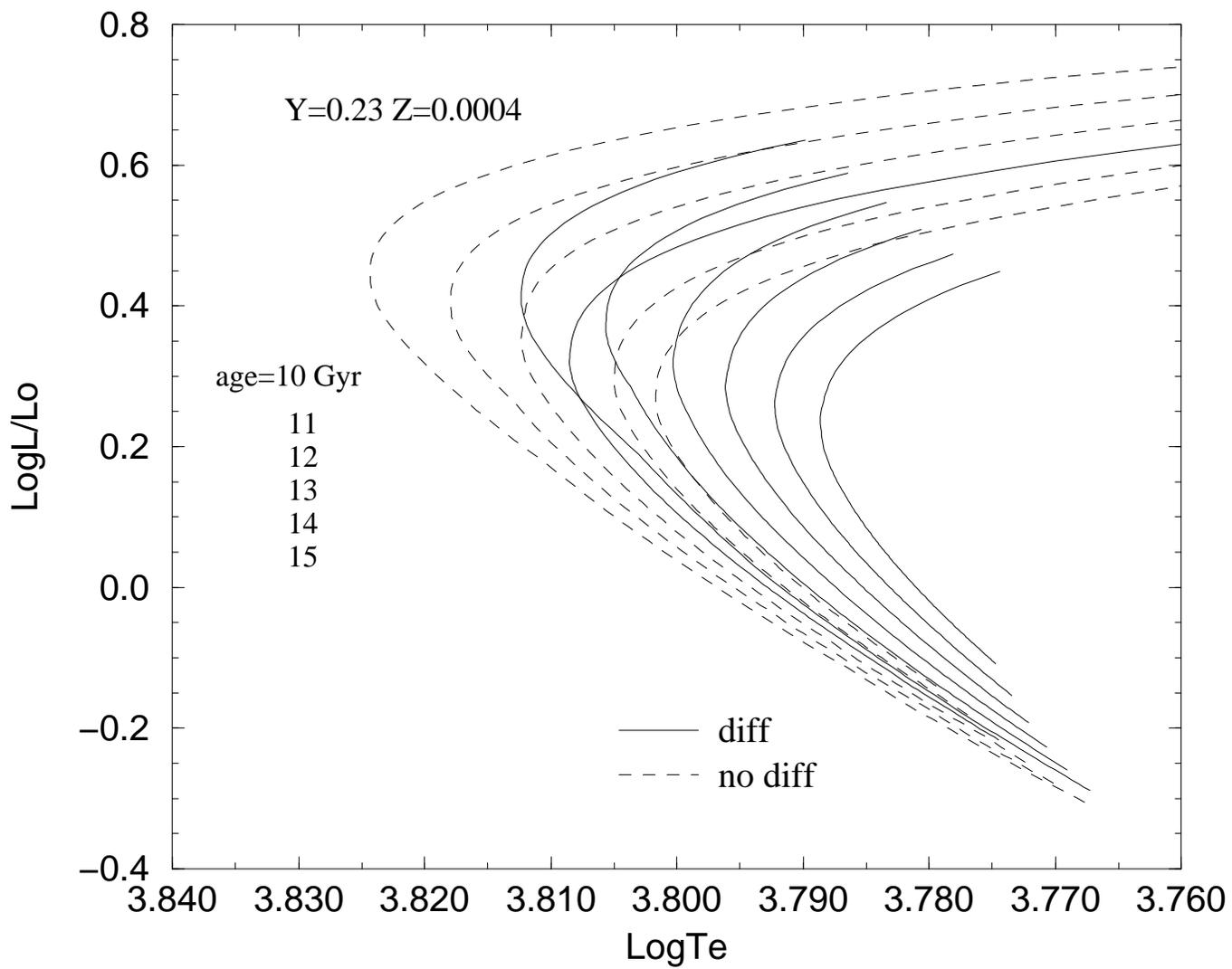
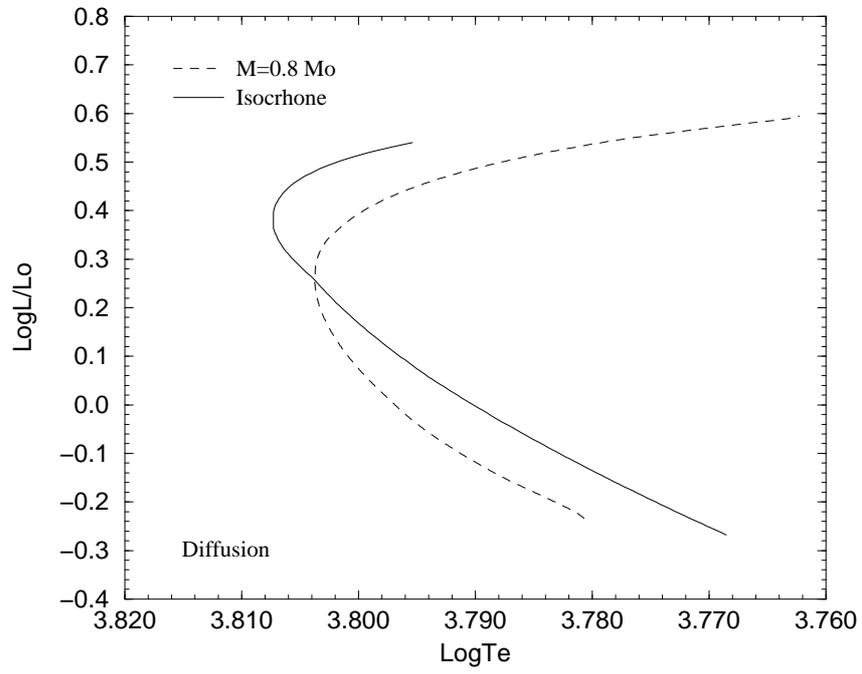
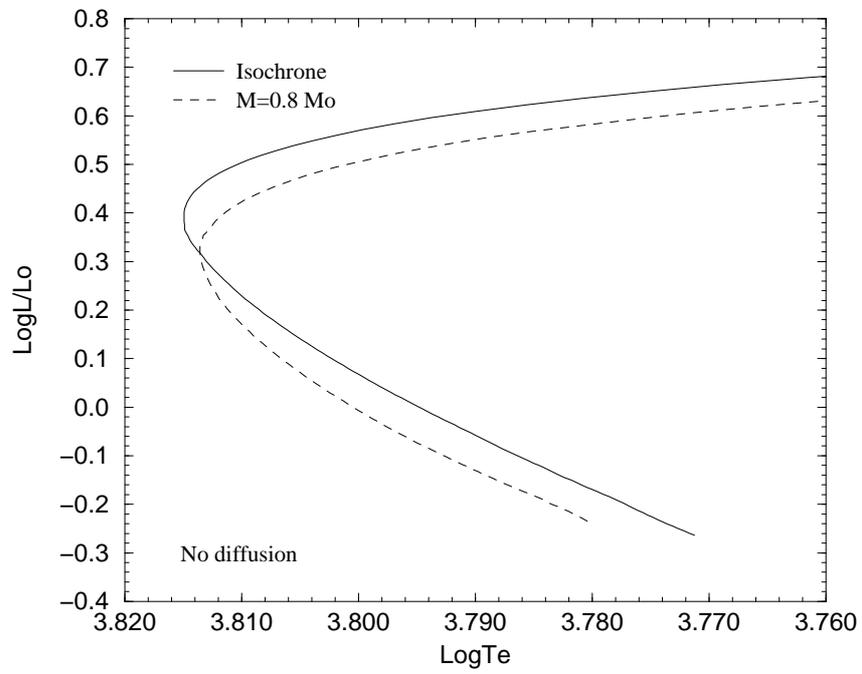
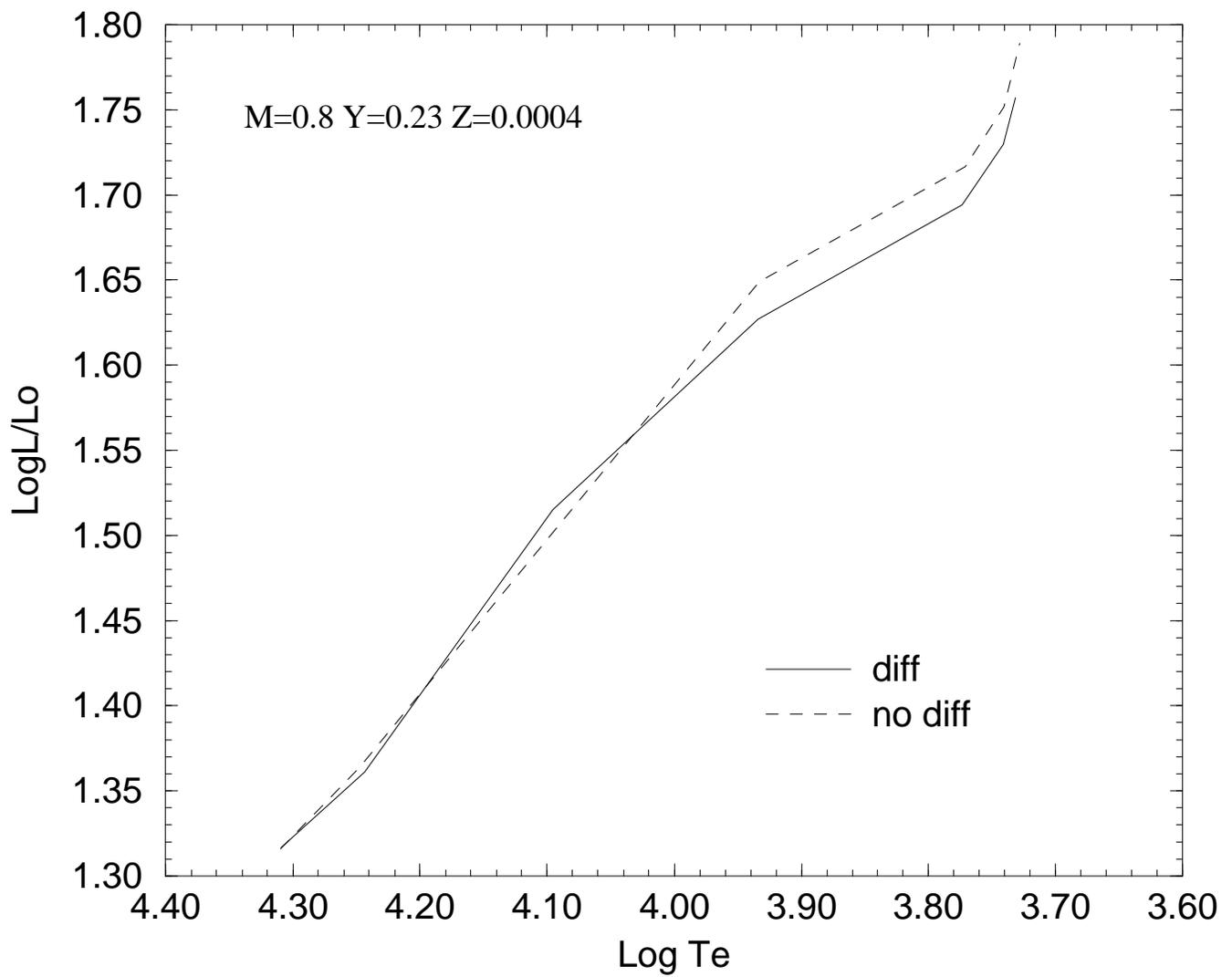
\bye